\documentclass[longauth]{aa} % for the long lists of affiliations 
\usepackage{graphicx,natbib}
\usepackage{txfonts}
\begin{document}
   \title{The GASP-WEBT monitoring of \object{3C 454.3} during the 2008 optical-to-radio and $\gamma$-ray outburst\thanks{The radio-to-optical data 
   presented in this paper are stored in the GASP-WEBT archive; for questions regarding their availability,
   please contact the WEBT President Massimo Villata ({\tt villata@oato.inaf.it}).}}

%   \subtitle{}

   \author{M.~Villata                 \inst{ 1}
   \and   C.~M.~Raiteri               \inst{ 1}
   \and   M.~A.~Gurwell               \inst{ 2}
   \and   V.~M.~Larionov              \inst{ 3,4,5}
   \and   O.~M.~Kurtanidze            \inst{ 6}
   \and   M.~F.~Aller                 \inst{ 7}
   \and   A.~L\"ahteenm\"aki          \inst{ 8}
   \and   W.~P.~Chen                  \inst{ 9}
   \and   K.~Nilsson                  \inst{10}
   \and   I.~Agudo                    \inst{11}
   \and   H.~D.~Aller                 \inst{ 7}
   \and   A.~A.~Arkharov              \inst{ 4}
   \and   U.~Bach                     \inst{12}
   \and   R.~Bachev                   \inst{13}
   \and   P.~Beltrame                 \inst{14}
   \and   E.~Ben\'{\i}tez             \inst{15}
   \and   C.~S.~Buemi                 \inst{16}
   \and   M.~B\"ottcher               \inst{17}
   \and   P.~Calcidese                \inst{18}
   \and   D.~Capezzali                \inst{19}
   \and   D.~Carosati                 \inst{19}
   \and   D.~Da~Rio                   \inst{14}
   \and   A.~Di~Paola                 \inst{20}
   \and   M.~Dolci                    \inst{21}
   \and   D.~Dultzin                  \inst{15}
   \and   E.~Forn\'e                  \inst{22}
   \and   J.~L.~G\'omez               \inst{11}
   \and   V.~A.~Hagen-Thorn           \inst{ 3,5}
   \and   A.~Halkola                  \inst{10}
   \and   J.~Heidt                    \inst{23}
   \and   D.~Hiriart                  \inst{15}
   \and   T.~Hovatta                  \inst{ 8}
   \and   H.-Y.~Hsiao                 \inst{ 9}
   \and   S.~G.~Jorstad               \inst{24}
   \and   G.~N.~Kimeridze             \inst{ 6}
   \and   T.~S.~Konstantinova         \inst{ 3}
   \and   E.~N.~Kopatskaya            \inst{ 3}
   \and   E.~Koptelova                \inst{ 9}
   \and   P.~Leto                     \inst{16}
   \and   R.~Ligustri                 \inst{14}
   \and   E.~Lindfors                 \inst{10}
   \and   J.~M.~Lopez                 \inst{15}
   \and   A.~P.~Marscher              \inst{24}
   \and   M.~Mommert                  \inst{23}
   \and   R.~Mujica                   \inst{25}
   \and   M.~G.~Nikolashvili          \inst{ 6}
   \and   N.~Palma                    \inst{17}
   \and   M.~Pasanen                  \inst{10}
   \and   M.~Roca-Sogorb              \inst{11}
   \and   J.~A.~Ros                   \inst{22}
   \and   P.~Roustazadeh              \inst{17}
   \and   A.~C.~Sadun                 \inst{26}
   \and   J.~Saino                    \inst{10}
   \and   L.~A.~Sigua                 \inst{ 6}
   \and   M.~Sorcia                   \inst{15}
   \and   L.~O.~Takalo                \inst{10}
   \and   M.~Tornikoski               \inst{ 8}
   \and   C.~Trigilio                 \inst{16}
   \and   R.~Turchetti                \inst{14}
   \and   G.~Umana                    \inst{16}
 }

   \offprints{M.\ Villata}

   \institute{
 % 1
          INAF, Osservatorio Astronomico di Torino, Italy                                                     
 %           2
   \and   Harvard-Smithsonian Center for Astrophysics, MA, USA                                                
 %           3
   \and   Astronomical Institute, St.-Petersburg State University, Russia                                     
 %           4
   \and   Pulkovo Observatory, Russia                                                                         
 %           5
   \and   Isaac Newton Institute of Chile, St.-Petersburg Branch, Russia                                      
 %           6
   \and   Abastumani Astrophysical Observatory, Georgia                                                       
 %           7
   \and   Department of Astronomy, University of Michigan, MI, USA                                            
 %           8
   \and   Mets\"ahovi Radio Observatory, Helsinki University of Technology TKK, Finland                       
 %           9
   \and   Institute of Astronomy, National Central University, Taiwan                                         
 %          10
   \and   Tuorla Observatory, Department of Physics and Astronomy, University of Turku, Finland               
 %          11
   \and   Instituto de Astrof\'{\i}sica de Andaluc\'{\i}a, CSIC, Spain                                        
 %          12
   \and   Max-Planck-Institut f\"ur Radioastronomie, Germany                                                  
 %          13
   \and   Institute of Astronomy, Bulgarian Academy of Sciences, Bulgaria                                     
 %          14
   \and   Circolo Astrofili Talmassons, Italy                                                                 
 %          15
   \and   Instituto de Astronom\'{\i}a, Universidad Nacional Aut\'onoma de M\'exico, Mexico                   
 %          16
   \and   INAF, Osservatorio Astrofisico di Catania, Italy                                                    
 %          17
   \and   Astrophysical Institute, Department of Physics and Astronomy, Ohio University, OH, USA              
 %          18
   \and   Osservatorio Astronomico della Regione Autonoma Valle d'Aosta, Italy                                
 %          19
   \and   Armenzano Astronomical Observatory, Italy                                                           
 %          20
   \and   INAF, Osservatorio Astronomico di Roma, Italy                                                       
 %          21
   \and   INAF, Osservatorio Astronomico di Collurania Teramo, Italy                                          
 %          22
   \and   Agrupaci\'o Astron\`omica de Sabadell, Spain                                                        
 %          23
   \and   ZAH, Landessternwarte Heidelberg-K\"onigstuhl, Germany                                              
 %          24
   \and   Institute for Astrophysical Research, Boston University, MA, USA                                    
 %          25
   \and   INAOE, Mexico                                                                                       
 %          26
   \and   Department of Physics, University of Colorado Denver, CO, USA                                       
 }

   \date{}

  \abstract
  % context heading (optional)
  % {} leave it empty if necessary  
   {Since 2001, the radio quasar 3C 454.3 has undergone a period of high optical activity, culminating in the brightest optical state ever observed, during the 2004--2005 outburst. The Whole Earth Blazar Telescope (WEBT) consortium has carried out several multifrequency campaigns to follow the source behaviour.} 
  % aims heading (mandatory)
   {The GLAST-AGILE Support Program (GASP) was born from the WEBT to provide 
long-term continuous optical-to-radio monitoring of a sample of $\gamma$-loud blazars, during the operation of the AGILE and GLAST (now known as Fermi GST) $\gamma$-ray satellites. The main aim is to shed light on the mechanisms producing the high-energy radiation, through correlation analysis with the low-energy emission. Thus, since 2008 the monitoring task on 3C 454.3 passed from the WEBT to the GASP, while both AGILE and Fermi detected strong $\gamma$-ray emission from the source.}
  % methods heading (mandatory)
   {We present the main results obtained by the GASP at optical, mm, and radio frequencies in the 2008--2009 season, and compare them with the WEBT results from previous years.}
  % results heading (mandatory)
   {An optical outburst was observed to peak in mid July 2008, when Fermi detected the brightest $\gamma$-ray levels. A contemporaneous mm outburst maintained its brightness for a longer time, until the cm emission also reached the maximum levels. The behaviour compared in the three bands suggests that the variable relative brightness of the different-frequency outbursts may be due to the changing orientation of a curved inhomogeneous jet.
The optical light curve is very well sampled during the entire season, which is also well covered by
the various AGILE and Fermi observing periods. The relevant cross-correlation studies will be very important in constraining high-energy emission models.}
  % conclusions heading (optional), leave it empty if necessary 
  {}

%   \titlerunning{}

   \maketitle
%
%________________________________________________________________

\section{Introduction}

   \begin{figure*}
   \sidecaption
   \includegraphics[width=13cm]{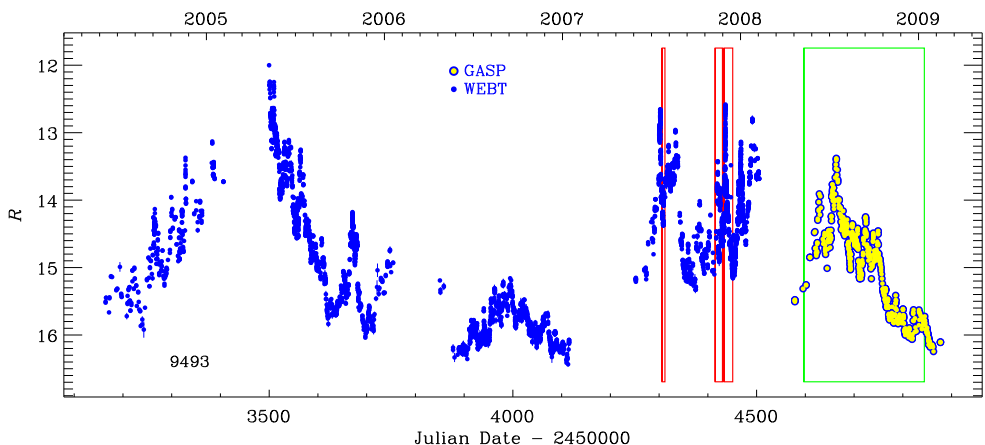}
%% PLEASE DON'T CHANGE THE FIGURE SIZE [width=13cm] AND ITS LOCATION IN THE TEXT, IF POSSIBLE
      \caption{$R$-band light curve of \object{3C 454.3} in the last five observing seasons, composed with data from the WEBT campaigns (blue dots) and from the GASP monitoring (yellow-filled circles); the boxes indicate various periods of $\gamma$-ray detection, as described in the text.}
      \label{fig1}
   \end{figure*}

Blazars (i.e.\ flat-spectrum radio quasars and BL Lacertae objects) constitute a specific class
of radio-loud active galactic nuclei whose highly variable emission is dominated by relativistically-beamed non-thermal radiation from a plasma jet. Blazars are detected at all wavelengths, from the radio to the $\gamma$-ray band. 
The low-energy non-thermal emission (from radio to optical, or sometimes to UV/X-rays) is due to synchrotron radiation, while the higher-energy emission is usually ascribed to inverse-Compton scattering of soft photons by 
the synchrotron-emitting relativistic electrons in the jet.
According to synchrotron-self-Compton (SSC) models, the soft photons are the synchrotron photons themselves, while in external-Compton (EC) models, seed photons come from outside the jet, in particular from the accretion disc or the broad line region. 
The different predictions on the multifrequency behaviour of the source from different models can be tested by the results of coordinated multiwavelength campaigns, coupling the high-energy data from satellite observations with ground-based radio-to-optical data.
In 1997, during the operation of the Compton Gamma Ray Observatory (CGRO, 1991--2000) satellite, this was one of the main motivations that led to the birth of the Whole Earth Blazar Telescope 
(WEBT)\footnote{{\tt http://www.oato.inaf.it/blazars/webt/}}, which involves a large number of telescopes at different longitudes, to obtain continuous monitoring during campaigns dedicated to single sources.
Ten years later, in 2007, following the launch of the $\gamma$-ray satellite
Astro-rivelatore Gamma a Immagini LEggero (AGILE), and in view of the anticipated launch of the Gamma-ray Large Area Space Telescope (GLAST, then renamed as the Fermi Gamma-ray Space Telescope),
the WEBT started a new project: the GLAST-AGILE Support Program (GASP; see e.g.\ \citealt{vil08}). Its primary aim is to provide 
long-term continuous optical-to-radio monitoring of a list of 28 $\gamma$-loud blazars, during the operation of these two satellites, 
by means of selected WEBT telescopes.

The flat-spectrum radio quasar \object{3C 454.3} ($z=0.859$) is one of the most studied blazars, especially after the observation of an exceptional bright state from the mm band to X-rays in May 2005.
That unprecedented outburst was monitored from the radio to the optical band by the WEBT, whose results were published by \citet{vil06} together with Chandra observations and VLBA imaging\footnote{Observations by the INTEGRAL and Swift satellites and by the REM telescope were presented in \citet{pia06}, \citet{gio06}, and \citet{fuh06}, respectively.}.
The WEBT continued to monitor the subsequent high radio activity; an interpretation of the optical-radio correlation was proposed by \citet{vil07}.
The subsequent observing seasons showed an alternation of quiescent and active states; the relevant WEBT monitoring and spectral energy distribution (SED) results were presented and analysed in \citet{rai07b,rai08b,rai08c} together with optical--UV and X-ray observations by the XMM-Newton and Swift satellites and spectroscopic monitoring in the near-IR. During the renewed optical activity of the second half of 2007, AGILE detected the source several times, at the brightest levels ever observed up to that time, as shown by \citet{ver08,ver09} and \citet{don09b}. In the last two papers, the authors also performed a cross-correlation analysis between the $\gamma$-ray fluxes in November and December 2007 and the corresponding WEBT optical data, showing a possible delay of the $\gamma$-ray emission of about 1 day or less. In particular, the detection of an exceptionally fast and strong flare in both the optical and $\gamma$-ray bands on December 12 seems to constrain the time lag to be less than 12 hours.

Thus, the WEBT followed the radio-to-optical behaviour of 3C 454.3 until February 2008; since then, the GASP continued this task in the last observing season 2008--2009, as part of its 28-source monitoring effort.
Between late May and late June 2008, \citet{don08atel}, \citet{vit08atel}, and \citet{gas08atel} reported on various episodes of $\gamma$-ray activity detected by AGILE. In July, the GASP observed a bright optical--NIR flare accompanied by mm and cm radio activity \citep{vil08atel}.
In the meantime, GLAST/Fermi-GST started its operation and the Large Area Telescope (LAT), turned on in late June, promptly detected high $\gamma$-ray activity, at unprecedented emission levels, especially during the optical flare \citep{tos08atel}. \citet{pit08atel} reported on the further observation by AGILE in late July. Following the optical-to-radio activity detected in July, the GASP continued to observe an increasing flux at high radio frequencies in the subsequent months, leading to the highest levels ever recorded at 43 GHz \citep{bac08atel}.

\citet{abd09} present the results of the first three months of Fermi-LAT observations of 3C 454.3, when the source showed unprecedentedly strong and highly-variable $\gamma$-ray emission, with a peak flux of $F_{E \, > \, 100\rm\,MeV}\approx 1.2 \, 10^{-5}\rm\,photons\,cm^{-2}\,s^{-1}$. Their light curve covers (with a quasi-daily sampling) the period July 1 -- October 5, and the similarities with the optical light curve presented in this letter are evident. However, this letter is focused on the multifrequency behaviour observed by the GASP in the 2008--2009 observing season, and any consideration and analysis on the optical-$\gamma$ correlation and its impact on theoretical models is deferred to subsequent papers in collaboration with the AGILE and Fermi teams. In Sect.\ 2 we present the optical-to-radio observations with a selection of light curves; in Sect.\ 3 we discuss the results.

\section{Optical-to-radio observations and results}

We calibrated the optical $R$-band magnitudes with respect to Stars 1, 2, 3, and 4 from the photometric sequence by \citet{rai98}.
Figure \ref{fig1} shows the $R$-band data collected by the GASP in the 2008--2009 observing season (from April 2008 to February 2009; yellow-filled circles), together with WEBT data during the previous four seasons, since the big 2004--2005 outburst (blue dots; data from \citealt{vil06,vil07,rai07b,rai08b,rai08c}).
In total, we assembled about 9500 data points. The participating GASP optical observatories in the 2008--2009 season were: Abastumani, Lulin, Armenzano, Crimean, Roque de los Muchachos (KVA), Belogradchik, San Pedro Martir, St.\ Petersburg, Kitt Peak (MDM), Teide (BRT), Sabadell, Talmassons, Calar Alto\footnote{Calar Alto data were acquired as part of the MAPCAT (Monitoring AGN with Polarimetry at the Calar Alto Telescopes) project.}, L'Ampolla, New Mexico Skies, Valle d'Aosta, and Tuorla. Near-IR $JHK$ data were taken at Campo Imperatore.
In the figure, the boxes indicate periods of $\gamma$-ray detection: the red ones are those by AGILE in 2007 \citep{ver08,ver09,don09b}, while the green one represents the total period covered by AGILE and/or Fermi in 2008 -- early 2009.

   \begin{figure}
   \resizebox{\hsize}{!}{\includegraphics{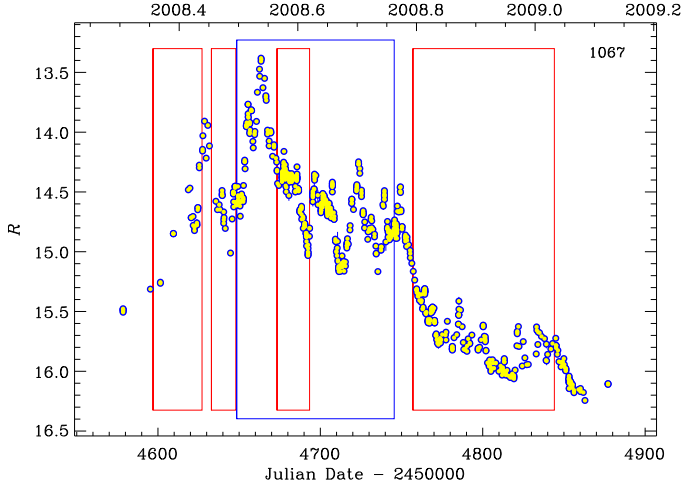}}
%% PLEASE DON'T CHANGE THE FIGURE SIZE (\resizebox{\hsize}{!}) AND ITS LOCATION IN THE TEXT, IF POSSIBLE
      \caption{Enlargement of the last season of the $R$-band light curve shown in Fig.\ \ref{fig1}, 
during the GASP monitoring; see text for explanation of the red and blue boxes.}
         \label{fig2}
   \end{figure}

   \begin{figure}
   \resizebox{\hsize}{!}{\includegraphics{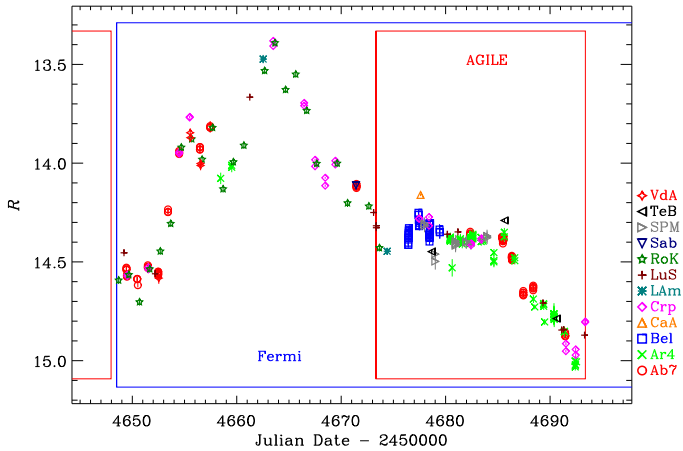}}
%% PLEASE DON'T CHANGE THE FIGURE SIZE (\resizebox{\hsize}{!}) AND ITS LOCATION IN THE TEXT, IF POSSIBLE
      \caption{Further enlargement of the $R$-band light curve around the brightest phase of the 2008 outburst, well monitored also by Fermi and AGILE; different symbols indicate different observatories.}
         \label{fig3}
   \end{figure}

   \begin{figure}
   \resizebox{\hsize}{!}{\includegraphics{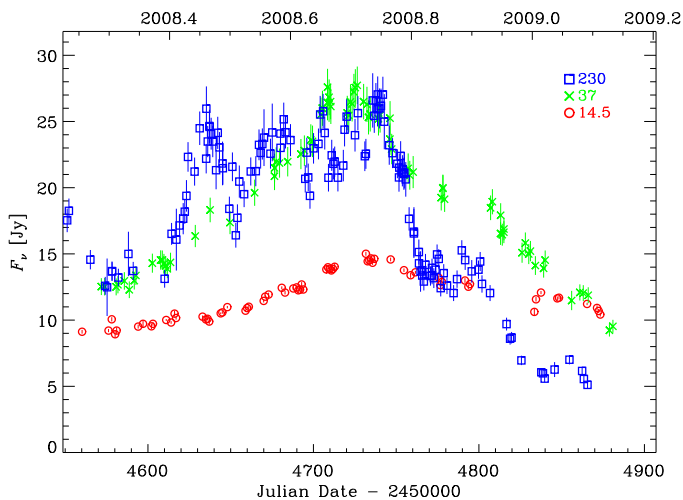}}
%% PLEASE DON'T CHANGE THE FIGURE SIZE (\resizebox{\hsize}{!}) AND ITS LOCATION IN THE TEXT, IF POSSIBLE
      \caption{GASP light curves of 3C 454.3 from late March 2008 to February 2009 at 230 GHz (blue squares), 37 GHz (green crosses), and 14.5 GHz (red circles).}
         \label{fig4}
   \end{figure}

The GASP light curve from April 2008 to February 2009 is better displayed in Fig.\ \ref{fig2}, where also the details of the $\gamma$-ray observing periods are given: red boxes are the AGILE ones, whereas the blue box indicates the July 1 -- October 5 period covered by the light curve in \citet{abd09}\footnote{The Fermi-LAT monitoring has continued beyond this period; preliminary light curves are shown at {\tt http://fermi.gsfc.nasa.gov/ssc/data/access/lat/msl\_lc/}.}. The optical flux started to increase in mid May and a first, noticeable flare peaked in mid June, at the end of the first AGILE period. A further, brighter flare was observed to double-peak on July 7--10 ($\rm JD\sim 2454655$--57), followed by the brightest phase of the outburst, peaking around July 16.1 ($\rm JD\sim 2454663.6$). This brightest part of the 2008 outburst is visible in more detail in Fig.\ \ref{fig3}, which reports the period from the start of the \citet{abd09} $\gamma$-ray light curve (July 1.0, $\rm JD=2454648.5$) until the end of the third 2008 AGILE period (August 14.88, $\rm JD=2454693.38$), with symbols differentiating the various observatories. Then the optical flux started to drop with a decreasing trend with several flares superposed, until the end of the season.

The GASP monitored the source also at mm and cm wavelengths: 345 and 230 GHz (SMA\footnote{These data were obtained as part of the normal monitoring program initiated by the SMA (see \citealt{gur07}).}), 43 GHz (Noto), 37 GHz (Mets\"ahovi), 22 GHz (Medicina), 14.5 GHz (UMRAO), 8 and 5 GHz (UMRAO and Medicina).
In Fig.\ \ref{fig4}, we show the best-sampled and most-significant light curves from late March 2008 to February 2009, at 230, 37, and 14.5 GHz\footnote{The other, less-sampled light curves show similar/intermediate trends.}.
The optical outburst was accompanied by very high activity at these lower frequencies; in particular, the 37 GHz data show levels never observed before, brighter than the previous historical maximum of early 2006. The mm--cm outbursts appear to be delayed with respect to the optical outburst, especially in the decreasing phase, with the delay increasing with wavelength.

Figure \ref{fig5} displays the flux behaviour at optical, mm, and radio frequencies in the last five seasons, starting from the 2004--2005 historical outburst, including the WEBT data from \citet{vil06,vil07} and \citet{rai07b,rai08b,rai08c}.
Compared to the previous optical activity, the 2008 outburst appears relatively modest, while the corresponding mm (230 GHz) outburst is much more spectacular, and the radio levels at 37 and 14.5 GHz are higher than in the previous events.

\section{Discussion}

   \begin{figure*}
   \sidecaption
   \includegraphics[width=13cm]{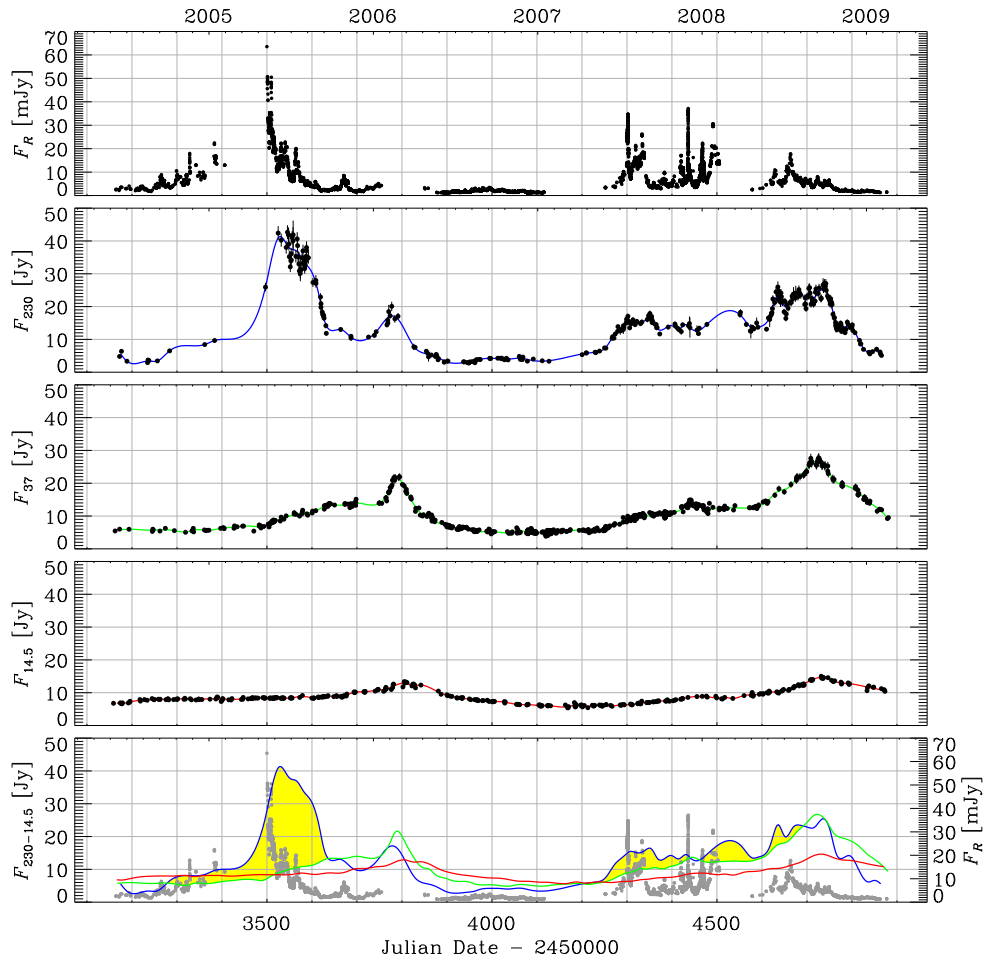}
%   \begin{figure}
%   \resizebox{\hsize}{!}{\includegraphics{2732fig5.eps}}
%% PLEASE DON'T CHANGE THE FIGURE SIZE (\resizebox{\hsize}{!}) AND ITS LOCATION IN THE TEXT, IF POSSIBLE
      \caption{$R$-band flux-density (de-reddened) light curve (top, mJy) and mm--cm light curves (Jy)
   at 230, 37, and 14.5 GHz from June 2004 to February 2009; in the bottom panel the optical data are reported together with the cubic spline interpolations through the 15-day binned data at 230 (blue line), 37 (green line), and 14.5 (red line) GHz; the regions where the 230 GHz spline exceeds the 37 GHz one are highlighted in yellow.}
         \label{fig5}
   \end{figure*}

The unprecedented optical outburst of 2005 was followed by an equally exceptional mm event, but, at lower frequencies, the signature of this event was only a moderate flux increase. According to \citet{vil07}, the different appearance of the outburst (which is caused by an emitting zone moving along the jet) at different frequencies was due to different viewing angles of the various emitting regions in the (curved) jet: smaller viewing angles yield a stronger Doppler enhancement of the flux. After that, an orientation change of the curved jet provided the extraordinary outburst at high radio frequencies of early 2006, as the counterpart of non-enhanced optical activity. This interpretation predicted a mild mm event, as one can now see in Fig.\ \ref{fig5} (see also \citealt{rai08c}). In other words, the ratios between the fluxes at different frequencies depend on the curved jet orientation. Following this picture, the strong (but not exceptional) optical activity in the 2007--2008 season, accompanied by mild mm and cm activity, would represent a geometrical configuration slightly favouring high frequencies. Then (2008--2009), the initial optical brightness is soon suppressed in favour of the mm one, which continues to be bright for a longer period, until the radio frequencies also brighten, more than previously, due to a particularly small viewing angle. It seems that the jet curvature can allow only two bands (e.g.\ optical and mm, or mm and cm) to be contemporaneously strongly Doppler enhanced: if one of the outer bands (i.e.\ optical and radio) is enhanced, the other is non-enhanced (see also the model for BL Lacertae in \citealt{vil09} and Fig.\ 4 therein). To make this evident, in the bottom panel of Fig.\ \ref{fig5} we plotted the optical data and cubic spline interpolations through the mm and cm light curves together. The yellow areas highlight the regions where the 230 GHz spline exceeds that at 37 GHz, i.e.\ when the mm emitting region is better aligned with the line of sight than the cm one. As expected, the larger the 230/37 GHz flux ratio, the higher the corresponding optical activity, and the optical appears always low when the 230/37 GHz ratio is less than 1. A deeper, quantitative analysis of this correlation is beyond the aim of this short letter.

We have reported on the optical-to-radio flux behaviour of 3C 454.3 in the 2008--2009 season, as observed by the GASP, and have compared these results with the WEBT light curves of recent years, finding consistency with the curved jet scenario depicted by \citet{vil07}.
In particular, we have assembled a detailed optical light curve during and around the various detection periods by AGILE and Fermi, which will be very useful for cross-correlation studies with the $\gamma$-ray emission.

\begin{acknowledgements}

We thank the referee, R.\ Hartman, for useful suggestions.
The Torino team acknowledges financial support by the Italian Space Agency through contract 
ASI-INAF I/088/06/0 for the Study of High-Energy Astrophysics. 
The Submillimeter Array is a joint project between the Smithsonian
Astrophysical Observatory and the Academia Sinica Institute of Astronomy and
Astrophysics and is funded by the Smithsonian Institution and the Academia
Sinica.
The St.\ Petersburg team acknowledges support from Russian Foundation for Basic Researches, grant 09-02-00092.
AZT-24 observations at Campo Imperatore are made within an agreement between Pulkovo, Rome and Teramo observatories.
This research has made use of data from the University of Michigan Radio Astronomy Observatory,
which is supported by the National Science Foundation and by funds from the University of Michigan.
The Mets\"ahovi team acknowledges the support from the Academy of Finland.
The research has been supported by the Taiwan National Science Council grant No.\ 96-2811-M-008-033.
This paper is partly based on observations carried out at the German-Spanish Calar Alto Observatory, which is jointly operated by the MPIA and the IAA-CSIC. Acquisition of the MAPCAT data is supported in part by the Spanish ``Ministerio de Ciencia e Innovaci\'on" through grant AYA2007-67626-C03-03.
This work is partly based on observations with the Medicina and Noto radio telescopes operated by INAF -- Istituto di Radioastronomia.
\end{acknowledgements}

\end{document}